\documentclass[letterpaper]{article} % DO NOT CHANGE THIS
\usepackage[]{aaai24}  % DO NOT CHANGE THIS
\usepackage{times}  % DO NOT CHANGE THIS
\usepackage{helvet}  % DO NOT CHANGE THIS
\usepackage{courier}  % DO NOT CHANGE THIS
\usepackage[hyphens]{url}  % DO NOT CHANGE THIS
\usepackage{graphicx} % DO NOT CHANGE THIS
\usepackage{natbib}  % DO NOT CHANGE THIS AND DO NOT ADD ANY OPTIONS TO IT
\usepackage{caption} % DO NOT CHANGE THIS AND DO NOT ADD ANY OPTIONS TO IT
\usepackage{algorithm}
\usepackage{algorithmic}
\usepackage[utf8]{inputenc} % allow utf-8 input
\usepackage[T1]{fontenc}    % use 8-bit T1 fonts
\usepackage{url}            % simple URL typesetting
\usepackage{booktabs}       % professional-quality tables
\usepackage{amsfonts}       % blackboard math symbols
\usepackage{nicefrac}       % compact symbols for 1/2, etc.
\usepackage{microtype}      % microtypography
\usepackage{cite}
\usepackage{amssymb}
\usepackage{graphicx}
\usepackage{xcolor}
\usepackage{amssymb}
\usepackage{diagbox}
\usepackage{amsmath}
\usepackage{siunitx}
\usepackage{amsthm}
\usepackage{comment}
\usepackage{bm}
\usepackage{multirow}
\usepackage{algorithm}
\usepackage{amssymb}
\usepackage{algorithmic}
\usepackage{array}
\usepackage{booktabs,caption}
\usepackage{amsmath}
\usepackage{enumitem}
\usepackage{natbib}
\usepackage{bbding}
\usepackage[flushleft]{threeparttable}

\usepackage{xspace}
\newcommand{\method}{\textsc{EEGFormer}\xspace}
\usepackage{newfloat}
\usepackage{listings}
\DeclareCaptionStyle{ruled}{labelfont=normalfont,labelsep=colon,strut=off} % DO NOT CHANGE THIS
\floatstyle{ruled}
\newfloat{listing}{tb}{lst}{}
\floatname{listing}{Listing}
\title{EEGFormer: Towards Transferable and Interpretable Large-Scale\\EEG Foundation Model}
\author{
    Yuqi Chen\textsuperscript{\rm 1}\footnote{A preprint version of an ongoing work, conducted during Yuqi's internship at Microsoft Research. Correspondence to Kan Ren.},
    Kan Ren\textsuperscript{\rm 2},
    Kaitao Song\textsuperscript{\rm 1},
    Yansen Wang\textsuperscript{\rm 1},\\
    Yifan Wang\textsuperscript{\rm 2},
    Dongsheng Li\textsuperscript{\rm 1},
    Lili Qiu\textsuperscript{\rm 1}
}

\affiliations{
\textsuperscript{\rm 1} Microsoft Research \quad
\textsuperscript{\rm 2} ShanghaiTech University \\
yansenwang@microsoft.com \quad
renkan@shanghaitech.edu.cn
}
\usepackage{bibentry}
\begin{document}

\maketitle

\begin{abstract}
Self-supervised learning has emerged as a highly effective approach in the fields of natural language processing and computer vision. 
It is also applicable to brain signals such as electroencephalography (EEG) data,
given the abundance of available unlabeled data that exist in a wide spectrum of real-world medical applications ranging from seizure detection to wave analysis. 
The existing works leveraging self-supervised learning on EEG modeling mainly focus on pretraining upon each individual dataset corresponding to a single downstream task, which cannot leverage the power of abundant data, and they may derive sub-optimal solutions with a lack of generalization.
Moreover, these methods rely on end-to-end model learning which is not easy for humans to understand.
In this paper, we present a novel EEG foundation model, namely \method, 
pretrained on large-scale compound EEG data.
The pretrained model cannot only learn universal representations on EEG signals with adaptable performance on various downstream tasks but also provide interpretable outcomes of the useful patterns within the data.
To validate the effectiveness of our model, we extensively evaluate it on various downstream tasks 
and assess the performance under different transfer settings. 
Furthermore, we demonstrate how the learned model exhibits transferable anomaly detection performance and provides valuable interpretability of the acquired patterns via self-supervised learning.
\end{abstract}

\section{Introduction}

Scalp electroencephalography (EEG) are physiological signal data that provide valuable insight into the human brain activities and has extensive applications in healthcare, e.g., disease diagnosis and medical monitoring~\cite{lawhern2018eegnet, tang2021self, tang2023modeling,li2023protecting}. 
Despite the ease of collecting EEG signals, comprehending and interpreting them often requires extensive expertise from medical professionals. 
To address this challenge, recent research has focused on leveraging self-supervised learning techniques to learn meaningful representations from EEG data~\cite{yi2023learning, wang2023brainbert, li2022multi}. 
These learned representations can then be fine-tuned for various downstream tasks, including seizure detection~\cite{tang2021self, tang2023modeling}, abnormal detection~\cite{darvishi2023amplifying}, emotion recognition~\cite{yi2023learning, ye2022hierarchical, song2021variational, li2021multi}, etc. 
However, these existing works focus on pretraining upon each individual dataset corresponding to a single downstream task and fail to leverage the power of abundant data.
In this paper, our primary interest lies in exploring the potential of self-supervised learning using abundant large-scale unlabeled data without human annotations.

Moreover, interpretability is a crucial concern when applying machine learning models to real-world applications~\cite{peng2022beit, ali2022xai, leung2022temporal}, particularly in the healthcare community~\cite{mendoza2023labeling, gulamali2023clinically}. Prior research~\cite{tang2021self, wang2023brainbert} has predominantly relied on end-to-end model learning, which poses challenges for human comprehension. Models that lack interpretability have the potential to yield unsafe and irrational outcomes, thereby increasing the risk of severe medical malpractice.

To address the above issues, we introduce \method as a solution for large-scale EEG pretraining. 
Our primary objective is to investigate a discrete representation learning approach~\cite{van2017neural, fortuin2018som, peng2022beit, esser2021taming} 
specifically designed for EEG pretraining.
We provide the evidence that the utilization of vector quantized Transformer \cite{c:22} model can learn universal representations on EEG signals with adaptable performance on various downstream tasks compared to the conventional mask reconstruction strategy~\cite{nie2022time}.
Furthermore, the learned codebook, along with the discrete indices provides interpretable outcomes of the useful patterns within the data.

\begin{figure*}[ht]
    \centering
    \includegraphics[width=0.95\textwidth]{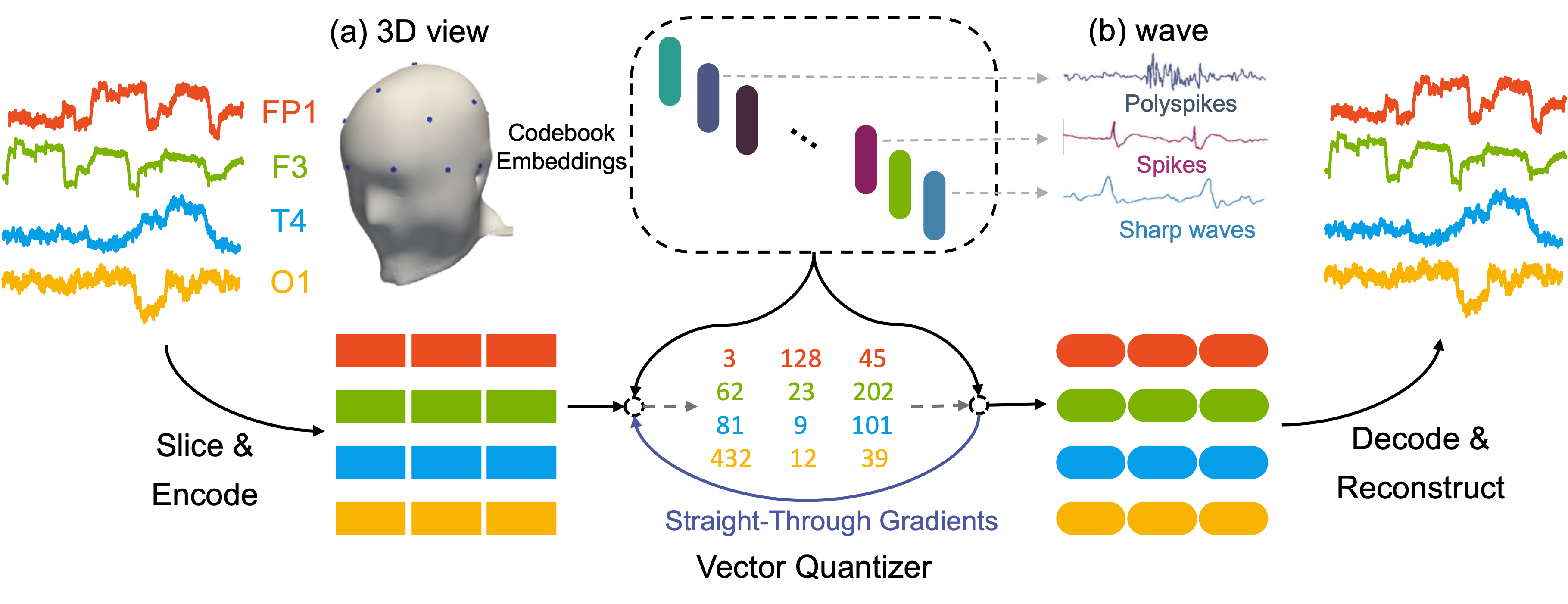}
    \caption{Overview of \method. Initially, multi-variate EEG signals are segmented into patches, which are then passed through a Transformer encoder. Subsequently, a vector-quantized model is employed to generate discrete indices. These indices are then fed into a shallow Transformer decoder.}
    \label{fig:model}
\end{figure*}

The contribution of the paper can be summarized as below:

\begin{itemize}[leftmargin=5mm]
    \item We propose a novel pretraining strategy for EEG data. \method adopts a discrete representation learning algorithm along with reconstruction loss. 
    \item We harness the plentiful EEG data available in the TUH Corpus \cite{harati2014tuh} to construct a foundational EEG model. This marks the pioneering effort in pretraining with a massive 1.7TB EEG dataset.
    \item We conduct a comprehensive analysis of the pretrained foundation model \method, evaluating its performance on four downstream corpora sourced from the TUH corpus. Additionally, we explore its transferability by applying it to the Neonate dataset~\cite{stevenson2019dataset} for neonatal seizure detection. 
    \item We provide an in-depth analysis of the learned codebook and demonstrate that the pretraining algorithm can provide transferable and interpretable representations.
\end{itemize}

\section{Related Work}

\paragraph{Pretraining for Time-Series Data}

Self-supervised learning for time-series data is a highly significant research hotspot.
% area of research. 
Many non-Transformer models have been developed to learn the representation of time series~\cite{franceschi2019unsupervised, tonekaboni2021unsupervised, yue2022ts2vec, eldele2021time}. 
Recently, ~\cite{nie2022time} introduced a Transformer-based approach that segments time series into patches, 
which leads to promising outcomes across various forecasting datasets. 
Furthermore, researchers are growing interested in utilizing pretrained large language models (LLMs) to enhance time series analysis~\cite{zhou2023one, gruver2023large}.
These methods are mainly on forecasting tasks and lack practical considerations of the model adaptation to different downstream tasks.

\paragraph{Pretraining for EEG data}

Electroencephalograms (EEGs) are widely employed for diagnosing neurological, and psychiatric, as well as in brain-machine interface applications. 
In the field of EEG signals, self-supervised learning has emerged as a promising approach~\cite{tang2021self, jiang2021self, kostas2021bendr}. 
SeqCLR ~\cite{mohsenvand2020contrastive} introduces a set of data augmentations for EEG and extends the SimCLR~\cite{chen2020simple} framework to extract channel-wise features on time-series EEG data. 
MMM~\cite{yi2023learning} focuses on spatial and topological modeling of EEG data and breaks the boundaries between different EEG topologies. 
However, these methods either apply self-supervision within the same dataset or test for a single downstream task, which cannot fully unleash the power of the self-supervised pretraining paradigm.
In this paper, 
our approach diverges the existing methods by leveraging the extensive multiple datasets of different tasks for pretraining purposes. 
Furthermore, we present a novel pretraining strategy that integrates discrete representation learning, thereby enhancing interpretability.

\section{\method: Vector-Quantized Pretraining Transformer for EEG Data}

This work aims to present a novel pretraining algorithm to derive a universal, reusable, and transferable EEG foundation model. 
In this paper, we focus on learning temporal patterns
among multi-channel EEG data. 
Specifically, we view EEG data as a multi-variate time series data, i.e., $X \in \mathbb{R}^{L \times C}$, where $L$ represents the length of the time series, and $C$ represents the number of channels (or variates)~\footnote{We mitigate the sample rate discrepancy by resampling the EEG data to a uniform rate of 250 Hz. Further, our analysis focuses on fixed-length 12-second EEG data following \cite{tang2021self}.}. 
Our primary goal is to develop a self-supervised learning algorithm that optimally leverages unlabelled data while enhancing interpretability. 
We introduce a customized vector-quantized pretraining approach designed for EEG data to accomplish this, as illustrated in Figure~\ref{fig:model}.
EEG signals can be encoded into discrete tokens, enabling interpretation through the analysis of these tokens, as is discussed in experiments. 
During the fine-tuning stage, 
the model and the 
codebook can be further fine-tuned to integrate specific domain-specific knowledge. 
In the subsequent subsections, we will provide a detailed description of the overall framework, including the preprocessing, EEG slicing, encoding module, decoding module, training algorithm, and fine-tuning processes.

\paragraph{Feature Preprocessing}

Converting EEG signals to the frequency domain is a common preprocessing technique. Following \cite{tang2021self}, given a time domain EEG signals, we perform fast Fourier transformation (FFT) to obtain frequency domain amplitude as input features.

\paragraph{Slice \& Encode}

To pretrain a time-series tokenizer, we first apply instance normalization to the frequency domain inputs. Then, we split each univariate time series into (non-)overlapped segments~\cite{nie2022time}. Specifically, for each variate (or channel), i.e., $x_c \in \mathbb{R}^{L}$ for the $c^\text{th}$ variate. 
Denote the patch length as $P$ and the stride as $S$,  the patching process will generate a sequence of patches $\mathrm{x}_c \in \mathbb{R}^{P \times N}$, where $N=\left(\lfloor \frac{L-P}{S} \rfloor + 2\right)$ indicates the number of patches. 
Given the input EEG data $x_c \in \mathbb{R}^{P \times N}$ for $c \in [1,.., C]$,
it is necessary to add position embedding before input to the Transformer encoder. Specifically, we map the dimension to $D$ via learnable weight matrix $\mathbf{w}_p \in \mathbb{R}^{P \times D}$ and adopt learnable position embedding, i.e., $\mathbf{w}_{pos} \in \mathbb{R}^{N \times D}$. 
Hence, the input vector is given by $\hat{x}_c = x_c^\top \mathbf{w}_p + \mathbf{w}_{pos}$. 
Finally, we forward $\hat{x}_c$ into a stack of Transformer encoder layers in a channel-independent manner~\cite{nie2022time}.

\paragraph{Vector Quantizer} 

The vector quantizer looks up the nearest neighbor in the codebook for each patch representation $h_i$. Let $\{v_1, v_2, \ldots , v_K\}$ denote the embeddings in the codebook. 
For the $i^\text{th}$ patch, its quantized code is calculated as $\boldsymbol{z}_i=\underset{j}{\arg \min }\left\|\boldsymbol{h}_i-\boldsymbol{v}_j\right\|_2$, where $j \in\{1,2, \cdots, K\}$. After quantizing the hidden vectors to discrete tokens, we feed the codebook embeddings $\left\{\boldsymbol{v}_{z_i}\right\}_{i=1}^N$ to the decoder model. 

\paragraph{Decode \& Reconstruct} 

The decoder model is a shallow Transformer model ~\cite{peng2022beit}. Upon passing through the decoder model, each variate generates an output denoted as $\hat{h}_c \in \mathbb{R}^{N \times D}$. We map the outputs to the same shape as the input through $\mathbf{w}_{o} \in \mathbb{R}^{D \times P}$ and $\mathbf{b}_{o} \in \mathbb{R}^{P}$, i.e., $x_{o}=\hat{h}_c\mathbf{w}_{o}+\mathbf{b}_{o}$. Finally, we reshape the output to match the shape of $X$, denoted as $X_{rec}$.

\paragraph{Training Loss}

The training objective of \method for each sample $X \in \mathcal{D}$ is to minimize
\begin{equation}
\resizebox{0.47\textwidth}{!}{$\|X_{rec} - X \|_2^2 + \sum_{i=1}^C \sum_{j=1}^{N} \|\operatorname{sg}\left[\boldsymbol{H}_{i,j} \right]-\boldsymbol{v}_{Z_{i,j}}\|_2^2-\left\| \boldsymbol{H}_{i,j}-\operatorname{sg}\left[\boldsymbol{v}_{Z_{i,j}}\right]\right\|_2^2 ~,$} 
\label{eq:loss}
\end{equation}
where $\mathrm{sg}[\cdot]$ stands for the stop-gradient operator which is an identity at the forward pass while having zero gradients during the backward pass~\cite{van2017neural}~\footnote{In Eq.~\eqref{eq:loss}, $\mathbf{H}$ denotes the hidden vectors for all the variates, whereas $\mathbf{h}$ stands for a single variate. Similarly for $Z$ and $z$.}.

\paragraph{Downstream Fine-Tuning}

To facilitate downstream fine-tuning, we utilize the pretrained model weights of both the encoder and the decoder modules. 
After obtaining the outputs $\hat{H} \in \mathbb{R}^{C \times N \times D}$, we feed them into the final layer for downstream tasks, such as classification or prediction.
Notably, as the codebook is amenable to fine-tuning, the training objective follows a formulation akin to that of Eq.~\eqref{eq:loss}.

\begin{table*}[t]
    \centering
    \caption{Experimental results on various downstream tasks. Within the table, $^{*}$ indicates a multi-classification task.}
    \label{tab:exp}
    \resizebox{0.9\textwidth}{!}{%
    \begin{tabular}{cccccccc}
    \toprule
        Model & Pretrain & Metric & TUAB & TUAR$^{*}$ & TUSL$^{*}$ & TUSZ & Neonate \\ 
    \midrule
        \multirow{2}{*}{EEGNet} & \multirow{2}{*}{\XSolidBrush} & (M-)AUROC& 0.841 ± .011 & 0.752 ± .006 & 0.635 ± .015 & 0.820 ± .030 & 0.793 ± .019   \\ 
        ~ & ~ & (M-)AUPRC & 0.832 ± .011 & 0.433 ± .025 & 0.351 ± .006 & 0.470 ± .017 & 0.499 ± .044 \\ 
    \midrule
        \multirow{2}{*}{TCN} & \multirow{2}{*}{\XSolidBrush} & (M-)AUROC & 0.841 ± .004 & 0.687 ± .011 & 0.545 ± .009 & 0.817 ± .004 & 0.731 ± .020   \\ 
        ~ & ~ & (M-)AUPRC & 0.831 ± .002 & 0.408 ± .009 & 0.344 ± .001 & 0.383 ± .010 & 0.398 ± .025 \\ 
    \midrule
        \multirow{2}{*}{EEG-GNN} & \multirow{2}{*}{\XSolidBrush} & (M-)AUROC & 0.840 ± .005 & 0.837 ± .022 & \textbf{0.721 ± .009} & 0.780 ± .006 & 0.760 ± .010   \\ 
        ~ & ~ & (M-)AUPRC & 0.832 ± .004 & \textbf{0.488 ± .015} & 0.381 ± .004 & 0.388 ± .023 & 0.419 ± .021 \\ 
    \midrule
        \multirow{2}{*}{GraphS4mer} & \multirow{2}{*}{\XSolidBrush} & (M-)AUROC & 0.864 ± .006 & 0.833 ± .006 & 0.632 ± .017 & 0.822 ± .034 & 0.719 ± .007   \\ 
        ~ & ~ & (M-)AUPRC & 0.862 ± .008 & 0.461 ± .024 & 0.359 ± .001 & 0.491 ± .001 & 0.374 ± .013 \\ 
    \midrule
    \midrule
        \multirow{2}{*}{BrainBERT} & \multirow{2}{*}{\Checkmark} & (M-)AUROC & 0.853 ± .002 & 0.753 ± .012 & 0.588 ± .013  & 0.814 ± .009 & 0.734 ± .019   \\ 
        ~ & ~ & (M-)AUPRC & 0.846 ± .003 & 0.350 ± .014 & 0.352 ± .003  & 0.386 ± .018 & 0.398 ± .027  \\ 
    \midrule
        \multirow{2}{*}{\method$_s$} & \multirow{2}{*}{\Checkmark} & (M-)AUROC & 0.862 ± .007 & 0.847 ± .013 & 0.683 ± .018 & 0.875 ± .004 & \textbf{0.842 ± .008} \\ 
        ~ & ~ & (M-)AUPRC & 0.862 ± .005 & \textbf{0.488 ± .012} & \textbf{0.397 ± .011} & 0.553 ± .014 & \textbf{0.578 ± .023}  \\ 
    \midrule
        \multirow{2}{*}{\method$_b$} & \multirow{2}{*}{\Checkmark} & (M-)AUROC & 0.865 ± .001 & 0.847 ± .014 & 0.713 ± .010 & 0.878 ± .006 & \textbf{0.842 ± .014} \\ 
        ~ & ~ & (M-)AUPRC & 0.867 ± .002 & 0.483 ± .026 & 0.393 ± .003 & \textbf{0.560 ± .010} & 0.568 ± .036  \\ 
    \midrule
        \multirow{2}{*}{\method$_l$} & \multirow{2}{*}{\Checkmark} & (M-)AUROC & \textbf{0.876 ± .003} & \textbf{0.852 ± .004} & 0.679 ± .013 & \textbf{0.883 ± .005} & 0.833 ± .017 \\ 
        ~ & ~ & (M-)AUPRC & \textbf{0.872 ± .001} & 0.483 ± .014 & 0.389 ± .003 & 0.556 ± .008 & 0.544 ± .026 \\ 
    \bottomrule
    \end{tabular}}
\end{table*}

\section{Experimental Results}

\paragraph{Datasets Description}
We pretrain our model on the Temple University EEG Corpus (TUH Corpus) \footnote{\url{https://isip.piconepress.com/projects/tuh_eeg/}}, which has collected over 1.7TB of unlabelled EEG data that are suitable for pretraining. 
We evaluate our model on five downstream datasets. i) TUAB corpus, which detects whether an EEG signal is normal or abnormal. ii) TUAR corpus, which contains annotations of 5 different artifacts. iii) TUSL corpus, which contains annotations of slowing events. v) TUSZ corpus, which contains annotations of seizure events. vi) Neonate dataset~\cite{stevenson2019dataset}, which contains annotation of neonatal seizures. Notably, the Neonate dataset is not a subset of the TUH dataset. 
Therefore, we consider the transferability of our pretraining strategy.

\paragraph{Parameter Setting}

We vary the encoder layers from $6$ to $12$, and the codebook size, i.e., $K$, from $512$ to $2048$. The decoder is a 3-layer Transformer. We set $D$ to $128$. Specifically, \method$_s$ adopts a 6-layer encoder and $K=512$, \method$_b$ adopts an 8-layer encoder and $K=1024$, and \method$_l$ adopts a 12-layer encoder and $K=2048$. 

\begin{figure}
    \centering
    \includegraphics[width=\linewidth]{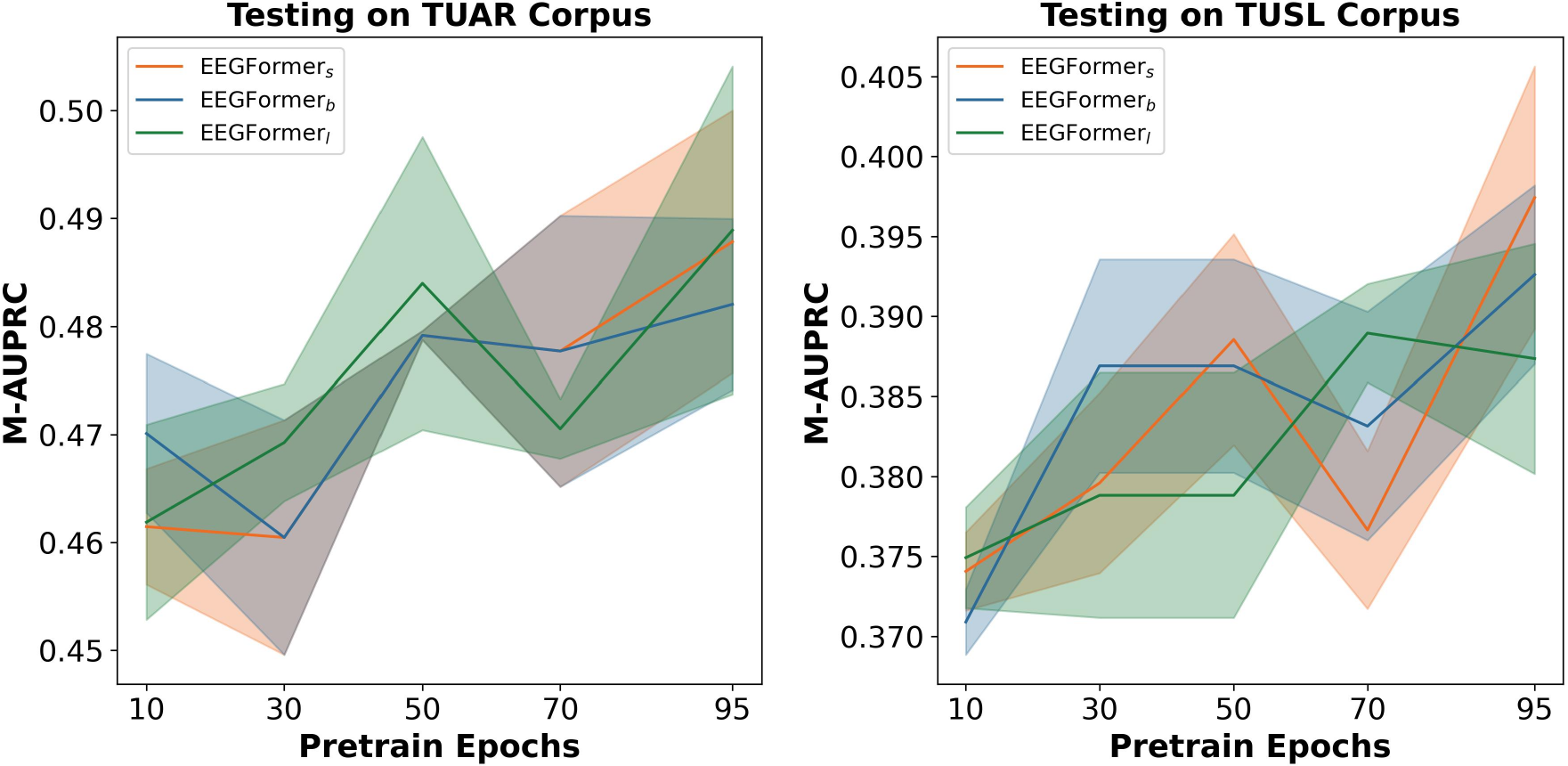}
    \caption{Influence of pretrain epochs on two TUH corpus.}
    \label{fig:epochs}
\end{figure}

\paragraph{Compared Baselines}

We compare \method with several baselines specifically for EEG data. i) EEGNet~\cite{lawhern2018eegnet} adopts a fully convolution network for EEG data. ii) TCN~\cite{bai2018empirical} adopts a dilated convolutional neural network. iii) EEG-GNN~\cite{tang2021self} adopts a graph neural network for capturing spatiotemporal dependencies
in EEGs. v) GraphS4mer~\cite{tang2023modeling} further adopts structured state space models or multivariate biosignals. Additionally, we also compare \method with self-supervised baselines. BrainBERT~\cite{wang2023brainbert} adopts neural signal processing techniques for producing superresolution time-frequency representations and pretrain with mask reconstruction loss\footnote{We use the pretrained weights from \url{https://drive.google.com/file/d/14ZBOafR7RJ4A6TsurOXjFVMXiVH6Kd_Q/view}.}.

\paragraph{Evaluation Metrics}

For detection tasks, we adopt the area under the receiver operating characteristic (AUROC) and the area under the precision-recall curve (AUPRC) for evaluation. For multi-classification tasks, we adopt macro AUROC (M-AUROC) and macro AUPRC (M-AUPRC) for evaluation.

\paragraph{Main Results}

The experimental results presented in Table ~\ref{tab:exp} clearly illustrate the effectiveness of our pretraining strategy in both in-dataset and transfer settings. Quantitatively, compared with the best baseline results, \method achieves a 15.8\% improvement on the Neonate dataset and a 14.1\% on the TUSZ under the AUPRC metric.

\paragraph{Influence of Pretrain Epochs}

We conducted experiments to examine the impact of pretraining epochs on various downstream corpora. The results of these experiments are illustrated in Figure \ref{fig:epochs}, Specifically, the results indicate that a longer pretraining period leads to notable enhancements in the performance of the downstream tasks.

\paragraph{Compared with Other Settings}

Table ~\ref{tab:lp} compares the performance of \method$_l$ using fine-tuning, linear probing, and supervising from scratch. By just fine-tuning the model head (linear probing), the performance of our model is already comparable with the supervised model (GraphS4mer). Additionally, we observe that the best results are observed with end-to-end fine-tuning.

\begin{table}[htbp]
    \centering
    \caption{Linear probe results on TUSL and TUAR corpus. Within the table, Sup stands for supervised learning from scratch, FT stands for self-supervised and fine-tuned, and LP stands for self-supervised and linear probing.}
    \label{tab:lp}
    \resizebox{\linewidth}{!}{%
    \begin{tabular}{cccccccc}
    \toprule
        Model & Type & Metric & TUAR & TUSL  \\ 
    \midrule
        
        \multirow{2}{*}{GraphS4mer} & \multirow{2}{*}{Sup} & M-AUROC &  0.833 ± .006 & 0.632 ± .017 \\ 
        ~ & ~ & M-AUPRC & 0.461 ± .024 & 0.359 ± .001 \\ 
    \midrule
        \multirow{2}{*}{\method$_l$} & \multirow{2}{*}{Sup} & M-AUROC & 0.822 ± .012 & 0.703 ± .033  \\ 
        ~ & ~ & M-AUPRC & 0.447 ± .015 & 0.374 ± .003  \\
    \midrule
        \multirow{2}{*}{\method$_l$} & \multirow{2}{*}{LP} & M-AUROC & 0.827 ± .000 & 0.657 ± .017  \\ 
        ~ & ~ & M-AUPRC & 0.469 ± .002 & 0.359 ± .003  \\
    \midrule
        \multirow{2}{*}{\method$_l$} & \multirow{2}{*}{FT} & M-AUROC & 0.852 ± .004 & 0.679 ± .013  \\ 
        ~ & ~ & M-AUPRC & 0.483 ± .014 & 0.389 ± .003  \\
    \bottomrule
    \end{tabular}}
\end{table}

\paragraph{Towards Seizure Localization}

\begin{figure}
    \centering
    \includegraphics[width=\linewidth]{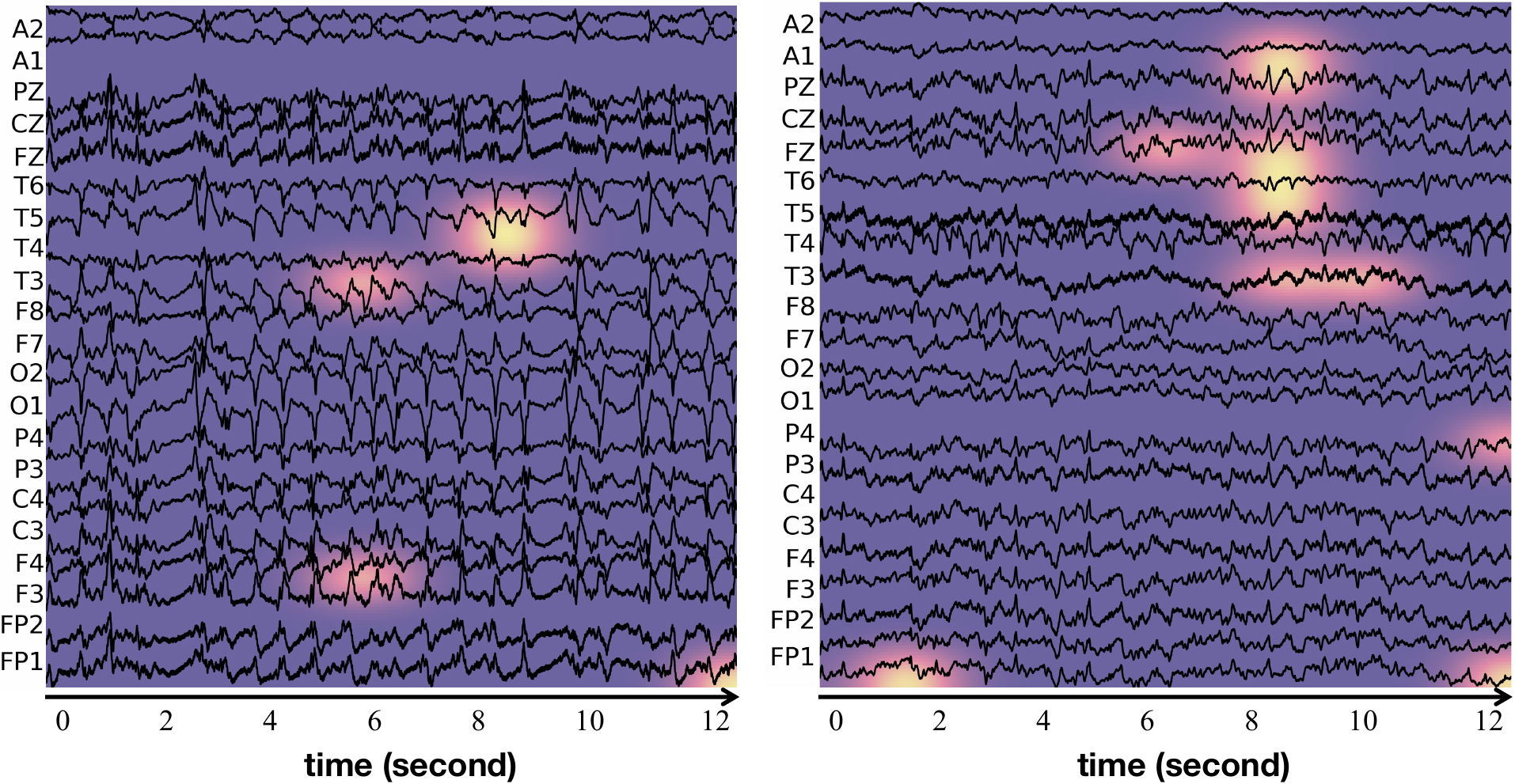}
    \caption{Interpretation results from naive Bayes model.}
    \label{fig:interpret}
\end{figure}

After the pertaining state, each EEG signal is discretized into multiple indices denoted as $I \in [1, ..., K]^{C \times N}$. To perform seizure detection in the TUSZ corpus using these pretrained indices, we first extract n-gram features for each data (e.g., 2-gram, 3-gram, and 4-gram). Next, we adopt a naive Bayes classifier based on n-gram features. Notably, we achieve an AUPRC of 0.292 and an AUROC of 0.741, without the need for fine-tuning the pretrained weight. Additionally, we extract the top-3 significant features with high posterior probability leading to seizure events, from the naive Bayes model. Figure \ref{fig:interpret} presents two cases, where the highlighted regions indicate the localization of seizures. It is worth noting that in the right figure, the highlighted segments correspond to the spike and slow wave complex in all the frontal lobe (Fz), parietal lobe (Pz), and temporal lobe (T3, T6), which indicates an epileptiform discharge (EPSP) followed by the refractory period of the affected neuron population after the large and synchronized neuron EPSP. This is often treated as one of the most important patterns for the diagnosis of epilepsy and the onset of a seizure event. Hence, these patterns are significant in enhancing the interpretability of the pretrained model.

\section{Conclusion}

In this paper, we introduce a novel EEG foundation model, named \method, for self-supervised learning using large-scale EEG data. Our approach leverages a vector-quantized learning algorithm to simultaneously learn a discrete codebook and representations of multi-variate EEG signals. We extensively evaluate our pretraining algorithm on different downstream tasks using the TUH dataset, demonstrating its effectiveness. 
Furthermore, we perform an analysis to showcase the interpretability of our pretraining model.

\bibliography{aaai24}

\end{document}